 \documentclass[12pt,preprint]{aastex}

\usepackage{graphicx}

\def \apj {ApJ}

\def \apjl {ApJ}
\def \solphys {Solar Phys.}

\newcommand{\citeN}[1]{\citeauthor{#1} (\citeyear{#1})}
\newcommand{\citeNP}[1]{\citeauthor{#1} \citeyear{#1}}

\shortauthors{Socas-Navarro}
\shorttitle{The 3D Structure of a Sunspot Magnetic Field}

\begin{document}

\title{The Three-Dimensional Structure of a Sunspot Magnetic Field}

\author{H. Socas-Navarro}
   	\affil{High Altitude Observatory, NCAR\thanks{The National Center
	for Atmospheric Research (NCAR) is sponsored by the National Science
	Foundation.}, 3450 Mitchell Lane, Boulder, CO 80307-3000, USA}
	\email{navarro@ucar.edu}

\date{}%

\begin{abstract}
Here we report on observations of the three-dimensional structure of a
sunspot magnetic field from the photosphere to the chromosphere, obtained
with the new visible/infrared spectro-polarimeter SPINOR. The observations,
interpreted with a non-LTE modeling technique, reveal a surprisingly complex
topology with areas of opposite-sign torsion, suggesting that flux-ropes of
opposite helicities may coexist together in the same spot.
\end{abstract}
   


The observations that we used in this work are first-light data from the new
instrument SPINOR (Spectro-Polarimeter for INfrared and Optical
Regions, \citeNP{SNEP+05a}). Still under development, SPINOR can already be used
for high-resolution full spectro-polarimetry at virtually any combination of
3 spectral regions in the 400 - 1000~nm range. The particular dataset that we
report on was acquired on 16 June 2004 at 15:16 UT. We observed two
photospheric Fe I lines (at 849.7 and 853.8 nm) and two chromospheric lines
of the Ca II infrared triplet (at 849.8 and 854.2 nm) in active region NOAA
0634 at a time of particularly good seeing. The spectrograph slit was scanned
over that region to construct a three-dimensional datacube, in such a way
that for each (x,y,$\lambda$) point we have the four Stokes parameters $I,Q,U$
and $V$. We made use of the new adaptive optics system (\citeNP{RHR+03}) of the
Dunn Solar Telescope. Combined with the excellent atmospheric conditions at
the time of the observations, we achieved a spatial resolution as good as
0.6" (note that this figure varies slightly in the scanning direction due to
temporal changes in the seeing conditions), which is among the best attained
thus far in this kind of observations.

\begin{figure*}
\epsscale{.80}
\plotone{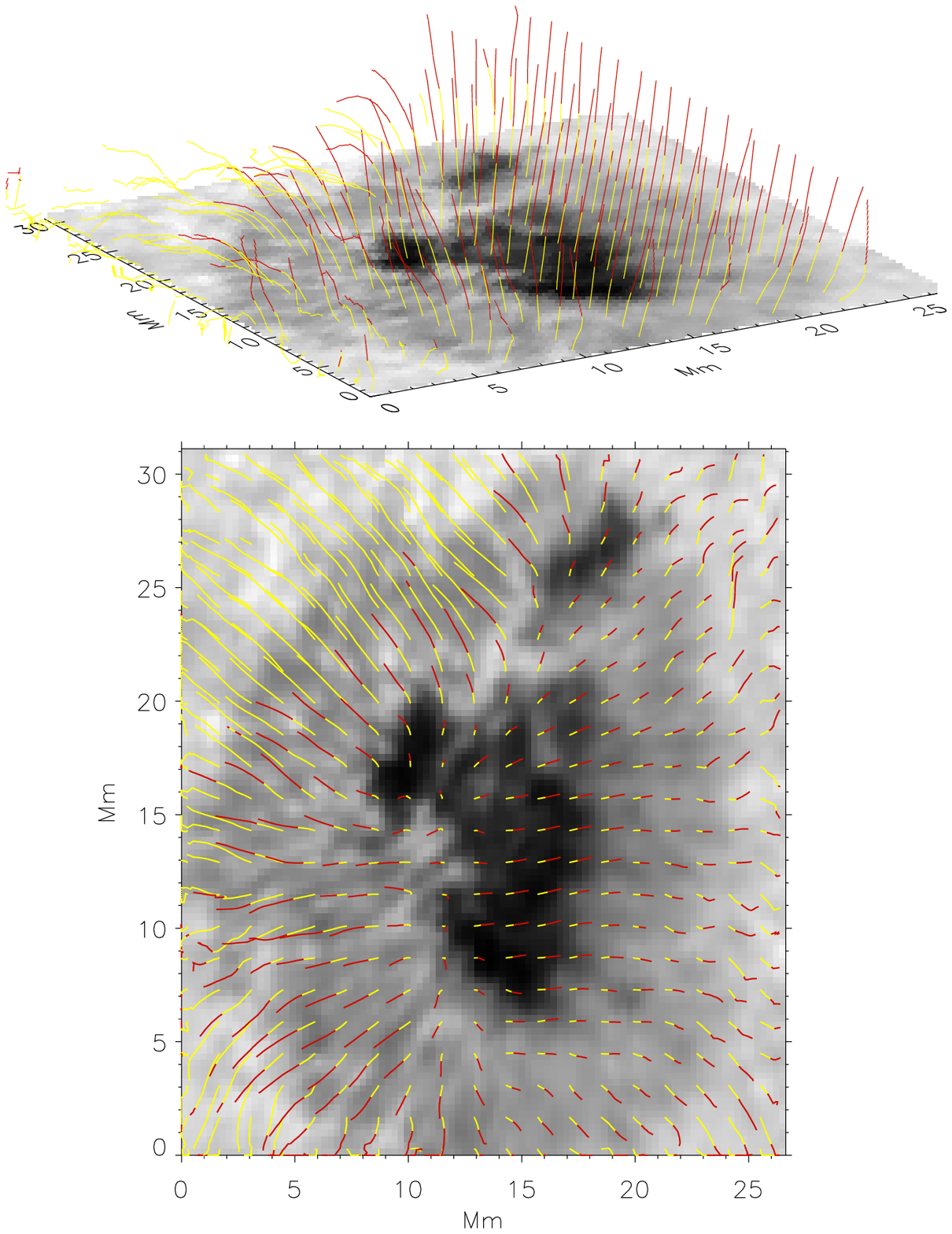}
\caption{ Three-dimensional structure of the magnetic field lines. Upper
  panel: Perspective view. Lower panel: Top view. The height range between 0
  and 800 km is represented in yellow, whereas the range between 800 and 1600
  km is in red. Axes are in megameters (Mm). Only field lines originating on
  a 5x5 (5x10) pixel grid are shown in the lower (upper) panel to avoid
  cluttering the figure. 
\label{fig:lines3D}
}
\end{figure*}

The sunspot subject to detailed analysis is rather irregular (see
Fig~\ref{fig:lines3D}) and exhibits two 
distinct umbral cores adjacent to the main umbra. One of them, above the main
umbra in the figure, is surrounded by its own penumbra. The other umbral core
is almost merged with the main umbra and is seen to its left separated by a
faint light bridge. 
The interpretation of the data was done using the Stokes inversion code
developed by \citeN{SNTBRC00a} for spectral lines formed in
non-LTE. The code infers the depth stratification of the temperature,
line-of-sight velocity and magnetic field vector that yields the best fit to
a particular set of Stokes spectra. 
The photospheric Fe blends in
the wings of the Ca lines are also computed by the code, providing a fully
connected picture of the whole atmosphere from the low photosphere to the
chromosphere.  

Each spatial point in the dataset is analyzed independently of the rest. In
order to ensure proper convergence and minimize the risk of the algorithm
settling in secondary minima, each inversion is repeated 10 times with
randomized initializations. The best solution is picked as representative of
the atmospheric conditions in the spatial location under study. The non-LTE
inversions are very computing-intensive. However, this kind of analysis is
necessary for accurate vector magnetometry (\citeNP{LMPS94}; \citeNP{SN02})
because radiative transfer 
and magneto-optical effects give rise to very complex dependences of the
observables on the atmospheric conditions.  
We employed a scheme by which the non-LTE inversion code is efficiently run
in parallel on several networked workstations. The hardware employed includes
three dedicated and three shared (mostly off-hours) Intel P4 processors
running Linux kernels. On average, we used the equivalent of 5 processors
running at a clockspeed of 2.7 GHz. The real-time employed in the entire
analysis was 29 days. This included a first inversion of a larger area (some
200x150 pixels) and a second pass at the area framed by a rectangle in Figure
1, of nearly 150x110 pixels. In the second pass, the profiles were averaged
over a 3x3 pixel (1.1"x0.66") box before inverting with the aim of improving
the signal-to-noise ratio. The noise level was thus reduced to approximately
1.5$\times$10$^{-4}$  in units of the quiet Sun continuum intensity. 

The detailed analysis outlined above produced a 3D reconstruction of the full
magnetic field vector in the active region. Figure~\ref{fig:lines3D} shows
the magnetic 
structure in the large sunspot, after deprojecting it to disk center and with
the magnetic field transformed to the solar reference frame. The 180-degree
ambiguity in the (observer frame) azimuth was resolved by picking the value
that results in a more radial field when converted to the solar coordinate
frame. As a consistency test, we computed the divergence of the field across
the entire map using boxes of 1.6~Mm in each dimension. It was found that the
divergence is always small compared to $B/l$ (the magnitude of the field
divided by the length of the box), with an average absolute value of
1.8\%. This argument, as well as the spatial coherence of the results obtained
(recall that each pixel was inverted independently), strongly supports the
reliability of our findings. 

\begin{figure*}
\epsscale{1.1}
\plotone{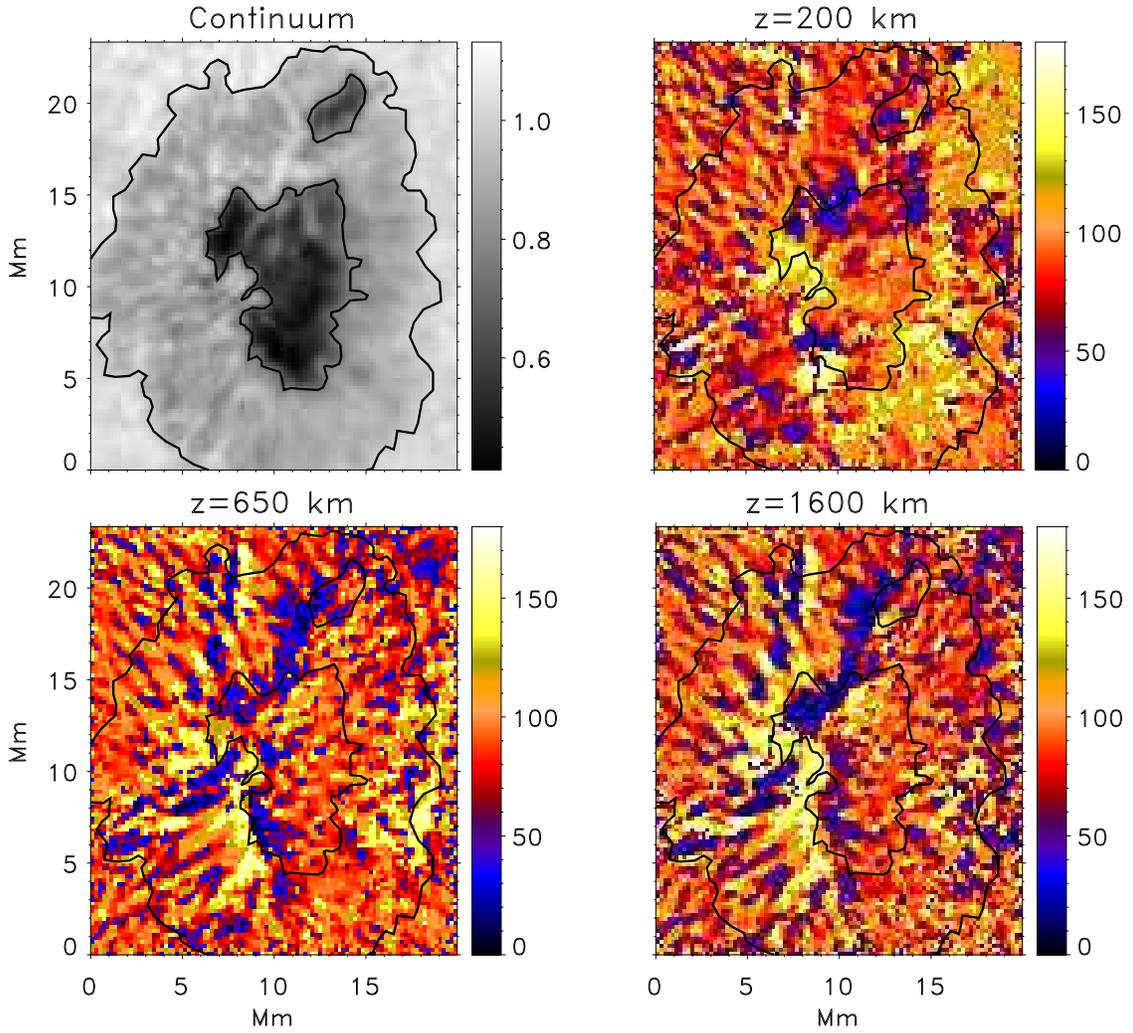}
\caption{Maps of the angle between vectors $\nabla \times \vec B$ and $\vec
  B$ at three different heights. The field is strictly force-free only in
  those pixels shown in dark blue (parallel) or light yellow (anti-parallel)
  colors. 
\label{fig:angles}
}
\end{figure*}

The east side of the sunspot (left of the figure) is reminiscent of what is
typically found in numerical simulations of a simple sunspot, with field
lines bending outwards in the penumbra. This area is partly force-free, as
can be deduced from Fig~\ref{fig:angles} (recall that the condition for a
force-free field is that $\nabla \times \vec B$ must be parallel or
anti-parallel to $\vec B$), particularly in the
chromosphere, but not entirely. The west side (right of the figure), on the
other hand, exhibits a very different structure and the field is mostly
non-force-free. Photospheric magnetograms of the region (not
show in the 
figures) reveal an intricate pattern of flux near the sunspot on that side,
suggesting a complex topology that would manifest itself also in the sunspot
structure, perturbing the field away from the classical picture of a
near-potential configuration. This idea is further supported by H$_{\alpha}$
images from the same day (e.g., Active Region Monitor at NASA Goddard Space
Flight Center's Solar Data Analysis Center:
http://www.solarmonitor.org/20040616/0634.html) showing considerable activity
on the west side of this sunspot.

\begin{figure*}
\epsscale{1.0}
\plotone{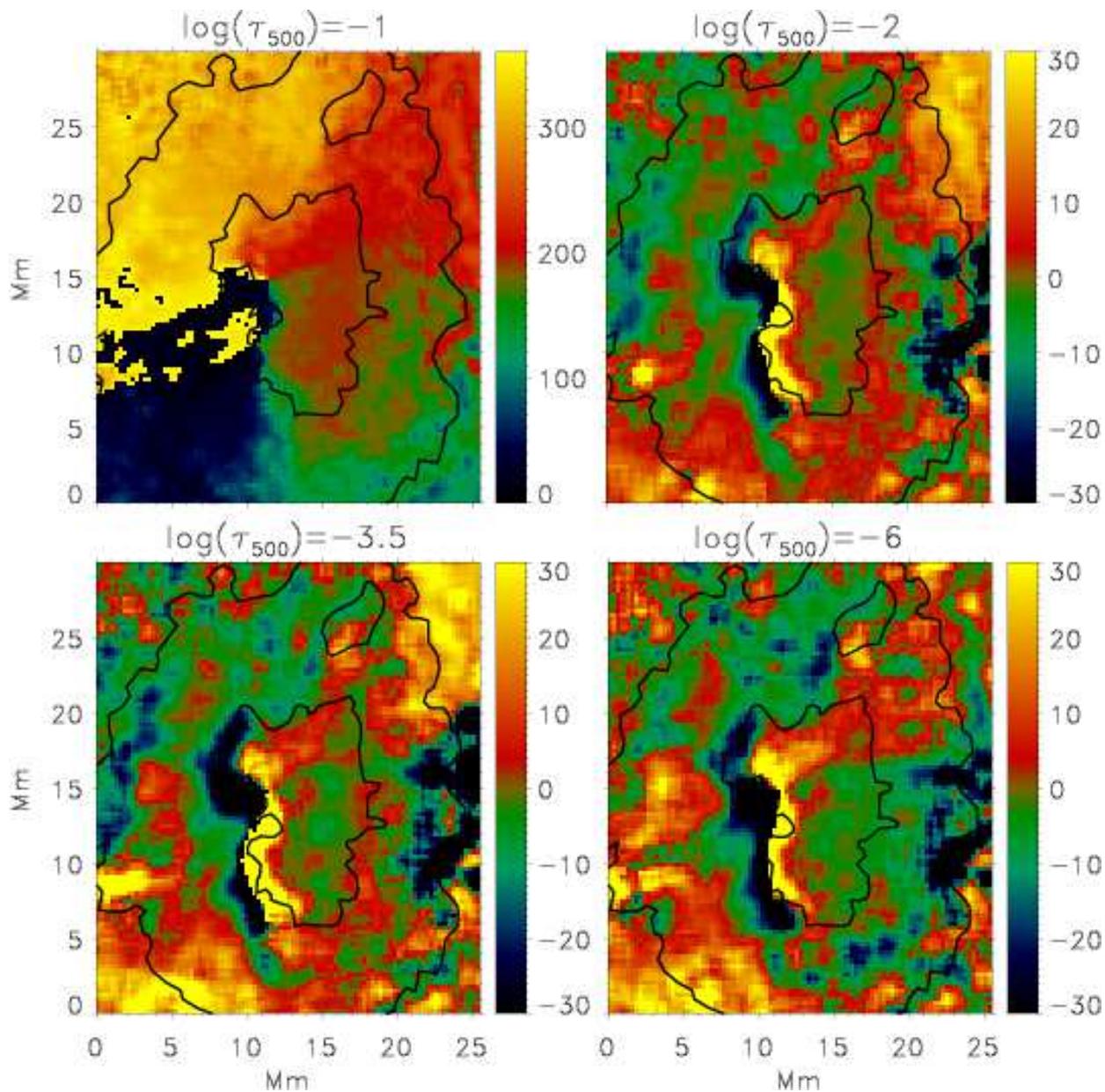}
\caption{Torsion maps. The color scale represents degrees measured
  counter-clockwise from the solar west direction. The upper-left panel shows
  the photospheric azimuth. All other panels show the difference between the
  azimuth at optical depth indicated by the label minus the photospheric
  azimuth (shown in the first panel), i.e. the magnetic torsion. The umbral
  and penumbral boundaries are overlaid for reference. 
\label{fig:torsion}
}
\end{figure*}

To avoid confusion with the terminology, which is sometimes used
with different meanings in the literature, we adopt the following definitions
in this work. We denote by "twist" the deviation of the field from the radial
direction and by "torsion" the vertical gradient of the azimuth\footnote{
The term azimuth here is defined as the angle between the (measured) magnetic 
field vector $\vec B$ and the solar East-West direction, measured
counter-clockwise from the solar West.  }
. In this
context, the concept of torsion is probably better defined from a
mathematical point of view because the radial direction is not well defined
in cases like the present one. Furthermore, while twist and torsion (as
defined here) refer essentially to the same concept in a rigid object, this
is not necessarily the case in non-rigid systems such as the magnetic field
lines. Imagine, for example, a magnetic field that does not vary along the
flux-rope axis but whose azimuth does not follow the radial direction. This
configuration would have twist but not torsion. 
A detailed representation of the magnetic torsion is provided in
Fig~\ref{fig:torsion}. This figure shows that the torsion is negative over
most of the sunspot, 
but there are also two large areas of positive torsion. The first one
corresponds roughly to the upper-right quadrant. It may be associated with
the direction connecting the main umbra to the umbral core located towards
the north-west. The torsion in this area decreases in magnitude as we move up
into the chromosphere. The other positive-torsion area is roughly the
lower-left quadrant and, contrarily to the first one, its magnitude increases
with height.  

The results presented in this paper reveal a very complex topology of sunspot
fields. The fact that most of the spot departs from the force-free regime is
quite surprising and has some important implications. While the force-free
approximation is a convenient way to estimate the magnetic field structure
(whether by means of extrapolations or using it to make empirical twist
determinations), our observations do not support its applicability in
general. 
Empirical determinations of twist or electric currents from 2D maps in
complex regions seem particularly questionable. In fact, such investigations
have even more caveats because one can only obtain one of the components of
the curl vector and also because the "measurement heights" at different
pixels have variations as large as 500 km in a sunspot. In summary, there are
no convenient shortcuts to measuring the actual 3D structure of the field via
full Stokes spectro-polarimetry and proper non-LTE modelling. Finally, it is
important to note that the non-force-freeness of the field provides potential
energy that may become available for chromospheric/coronal heating by means
of magnetic reconnection into a lower energy state.  The coexistence of
opposite-sign torsions is a rather surprising finding since, to the author's
best knowledge, it has not been predicted or proposed in any previous
theoretical models.



The physical mechanism that twists the field is yet to be established. Three
different processes have been proposed: a)the dynamo itself generates twisted
fields (\citeNP{CCN04}); b)Coriolis forces twist the flux-tubes during their
ascent (\citeNP{FFL+00}; \citeNP{FG00}); c)turbulent convective buffeting
twists the flux-tubes during their travel through the convection zone
(\citeNP{LFP98}). The mixed-twist scenario of our observations seems to be
difficult to explain by the first two possibilities, which should produce
flux-ropes of the same sign twist in a given active region. The convective
buffeting scenario, on the other hand, is essentially a turbulent process and
would naturally produce either sign twist in a more or less stochastic
manner. Therefore, while the three mechanisms are probably contributing with
different relative importance, our data indicates that the convective
buffeting is probably the dominant one. However, convective buffetting
cannot be the only mechanism for producing twist, since
the statistics of active regions clearly indicate that
there is also a systematic trend in the twist (see, for
example, \citeNP{PCM95}).


\clearpage

\clearpage

\clearpage

\end{document}